\documentclass[twocolumn,showpacs,showkeys,superscriptaddress,aps,floatfix]{revtex4}
\usepackage{amsmath,amssymb,booktabs,verbatim}
\usepackage{graphicx}\newcommand{\arxiv}[1]{\href{http://arxiv.org/abs/#1}{\texttt{arXiv:#1}}}
\usepackage[colorlinks=true,citecolor=black,filecolor=black,linkcolor=black,urlcolor=black,pdftex]{hyperref}
\newcommand{\1}[1]{\, \mathrm{#1}} 
\newcommand{\n}[1]{\mathrm{#1}}    
\newcommand{\percent}{\%}

\newcommand{\mpi}{\affiliation{Max-Planck-Institut f\"ur Physik, F\"ohringer Ring 6, D-80805 M\"unchen, Germany}}
\newcommand{\tum}{\affiliation{Physik-Department E15, Technische Universit\"at M\"unchen, D-85747 Garching, Germany}}
\newcommand{\oxford}{\affiliation{Department of Physics, University of Oxford, Oxford OX1 3RH, United Kingdom}}
\newcommand{\tubingen}{\affiliation{Eberhard-Karls-Universit\"at T\"ubingen, D-72076 T\"ubingen, Germany}}
\newcommand{\lngs}{\affiliation{INFN, Laboratori Nazionali del Gran Sasso, I-67010 Assergi, Italy}}
\newcommand{\coimbra}{\affiliation{on leave from: Departamento de Fisica, Universidade de Coimbra, P3004 516 Coimbra, Portugal}}
\newcommand{\deceased}{\affiliation{Deceased}}

\begin{document}

\title{Discrimination of Recoil Backgrounds in Scintillating Calorimeters}

\author{R.~F.~Lang}\email{rafael.lang@mpp.mpg.de}\mpi
\author{G.~Angloher}\mpi
\author{M.~Bauer}\tubingen
\author{I.~Bavykina}\mpi
\author{A.~Bento}\mpi \coimbra
\author{A.~Brown}\oxford
\author{C.~Bucci}\lngs
\author{C.~Ciemniak}\tum
\author{C.~Coppi}\tum
\author{G.~Deuter}\tubingen
\author{F.~von~Feilitzsch}\tum
\author{D.~Hauff}\mpi
\author{S.~Henry}\oxford
\author{P.~Huff}\mpi
\author{J.~Imber}\oxford
\author{S.~Ingleby}\oxford
\author{C.~Isaila}\tum
\author{J.~Jochum}\tubingen
\author{M.~Kiefer}\mpi
\author{M.~Kimmerle}\tubingen
\author{H.~Kraus}\oxford
\author{J.-C.~Lanfranchi}\tum
\author{M.~Malek}\oxford
\author{R.~McGowan}\oxford
\author{V.~B.~Mikhailik}\oxford
\author{E.~Pantic}\mpi
\author{F.~Petricca}\mpi
\author{S.~Pfister}\tum
\author{W.~Potzel}\tum
\author{F.~Pr\"obst}\mpi
\author{S.~Roth}\tum
\author{K.~Rottler}\tubingen
\author{C.~Sailer}\tubingen
\author{K.~Sch\"affner}\mpi
\author{J.~Schmaler}\mpi
\author{S.~Scholl}\tubingen
\author{W.~Seidel}\mpi
\author{L.~Stodolsky}\mpi
\author{A.~J.~B.~Tolhurst}\oxford
\author{I.~Usherov}\tubingen
\author{W.~Westphal}\tum\deceased

\date{\today}

\begin{abstract}
The alpha decay of $\n{{}^{210}Po}$ is a dangerous background to rare event searches. Here, we describe observations related to this alpha decay in the Cryogenic Rare Event Search with Superconducting Thermometers (CRESST). We find that lead nuclei show a scintillation light yield in our $\n{CaWO_4}$ crystals of $0.0142\pm0.0013$ relative to electrons of the same energy. We describe a way to discriminate this source of nuclear recoil background by means of a scintillating foil, and demonstrate its effectiveness. This leads to an observable difference in the pulse shape of the light detector, which can be used to tag these events. Differences in pulse shape of the phonon detector between lead and electron recoils are also extracted, opening the window to future additional background suppression techniques based on pulse shape discrimination in such experiments.
\end{abstract}

\maketitle

\section{Introduction}

Experiments that search for rare processes such as neutrinoless double-beta decay~\cite{avignone2005} or elastic scattering of Dark Matter particles~\cite{jungman1996} require a strong reduction of ambient radioactive backgrounds. One important source of radioactivity is radon gas, a member of the natural decay chain of $\n{{}^{238}U}$~\cite{firestone1996}. $\n{{}^{222}Rn}$ decays under emission of an alpha particle with a half-life of 3.8 days. This is long lived enough to plate out on materials in the vicinity of the detectors. In these alpha decays, recoiling nuclei can be implanted in surfaces. Eventually, the daughter nucleus $\n{{}^{210}Po}$ alpha decays into $\n{{}^{206}Pb}$ with a half-life of 138 days and a total energy of $5407\1{keV}$ available in the decay. On the one hand, this results in the ejection of a $5304\1{keV}$ alpha particle which can be a background to neutrinoless double-beta decay searches. On the other hand, the lead nucleus recoils with an energy of $103\1{keV}$. This can be a dangerous background in Dark Matter searches that needs to be dealt with, as we describe in the following.

\section{CRESST-II Detector Module}

For the CRESST-II experiment, we use scintillating $\n{CaWO_4}$ crystals as a target for a Dark Matter search. The crystals are cut in cylindrical shape of $4\1{cm}$ diameter and similar height, and weigh about $300\1{g}$ each. We expect Dark Matter to induce tungsten recoils in the target with energies below a few tens of keV. To enable a calorimetric measurement of these low energies, the crystals are cooled to $\sim 15\1{mK}$ where heat capacities are low. Tungsten films, evaporated on the crystals, are thermally stabilized in their transition to the superconducting state. This allows measurement of the rising film resistance caused by the temperature rise following a particle interaction in the crystal.

$\n{CaWO_4}$ is a good scintillator even at the cryogenic temperatures relevant here~\cite{mikhailik2007}. To measure the scintillation light following a particle interaction, a separate light detector is used. This allows discrimination of the dominant electron and gamma background from nuclear recoils which are expected from Dark Matter interactions, since the light output of these backgrounds is significantly higher than that of nuclear recoils. The light detector consists of a light absorbing wafer and another tungsten thermometer evaporated onto it. The scintillation light is absorbed in the wafer and results in a small but measurable temperature rise of the thermometer. Only about one percent of the total energy of a particle interaction is detected by the light channel~\cite{westphal2006,frank2002b,distefano2003}.

\begin{figure}[htbp]
\begin{center}\includegraphics[width=1\columnwidth]{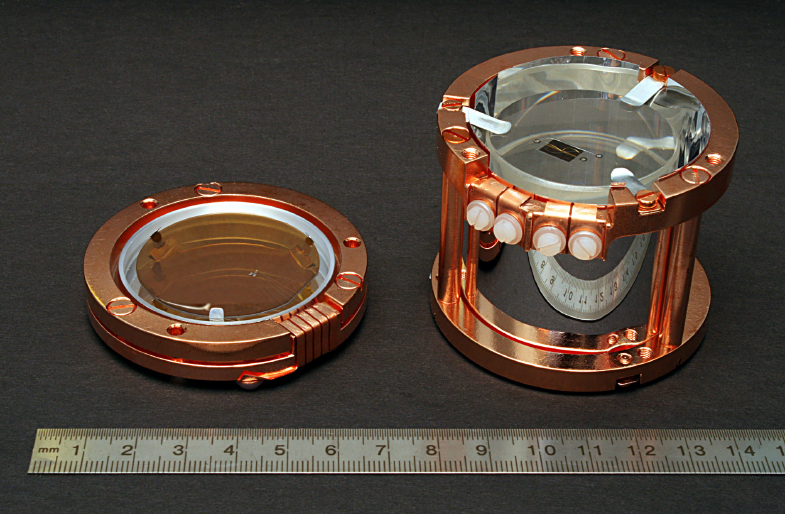}\end{center}
\vspace{-5mm}\caption{Picture of an opened CRESST-II detector module. On the right, the $\n{CaWO_4}$ crystal can be seen in its reflective housing, and with the thermometer evaporated on top of it. On the left, the light absorbing wafer is shown (silicon on sapphire) with its thermometer barely visible.}
\label{fig:modulfotok}\end{figure}

Figure~\ref{fig:modulfotok} shows one such detector module. It consists of the target crystal with the tungsten thermometer, which we refer to as the phonon detector, and the light absorbing wafer with its thermometer, the light detector. To improve the light collection, both detectors are enclosed in a reflective housing. 

To obtain the energy from the observed pulses, we use a template fit procedure. A template pulse is built from a large set of pulses with similar energy. This template is then fitted to each observed pulse, with three free parameters: the level of the baseline, the onset of the pulse, and its amplitude as the measure of interaction energy. Energy calibration is performed in a dedicated calibration run. Heater structures on the thermometer are used to inject large pulses in order to drive the thermometer into the normal conducting state. The thermometer response is then used to stabilize the operating temperature of the detector over many weeks. In addition, smaller pulses with fixed energies can also be injected to calibrate the thermometer at low energies where no external calibration source is available. Details of these procedures are laid out in~\cite{angloher2005,angloher2009}.

\section{Po-210 Decay Observed with Silver Reflectors}

\begin{figure}[htbp]
\begin{center}\includegraphics[width=1\columnwidth]{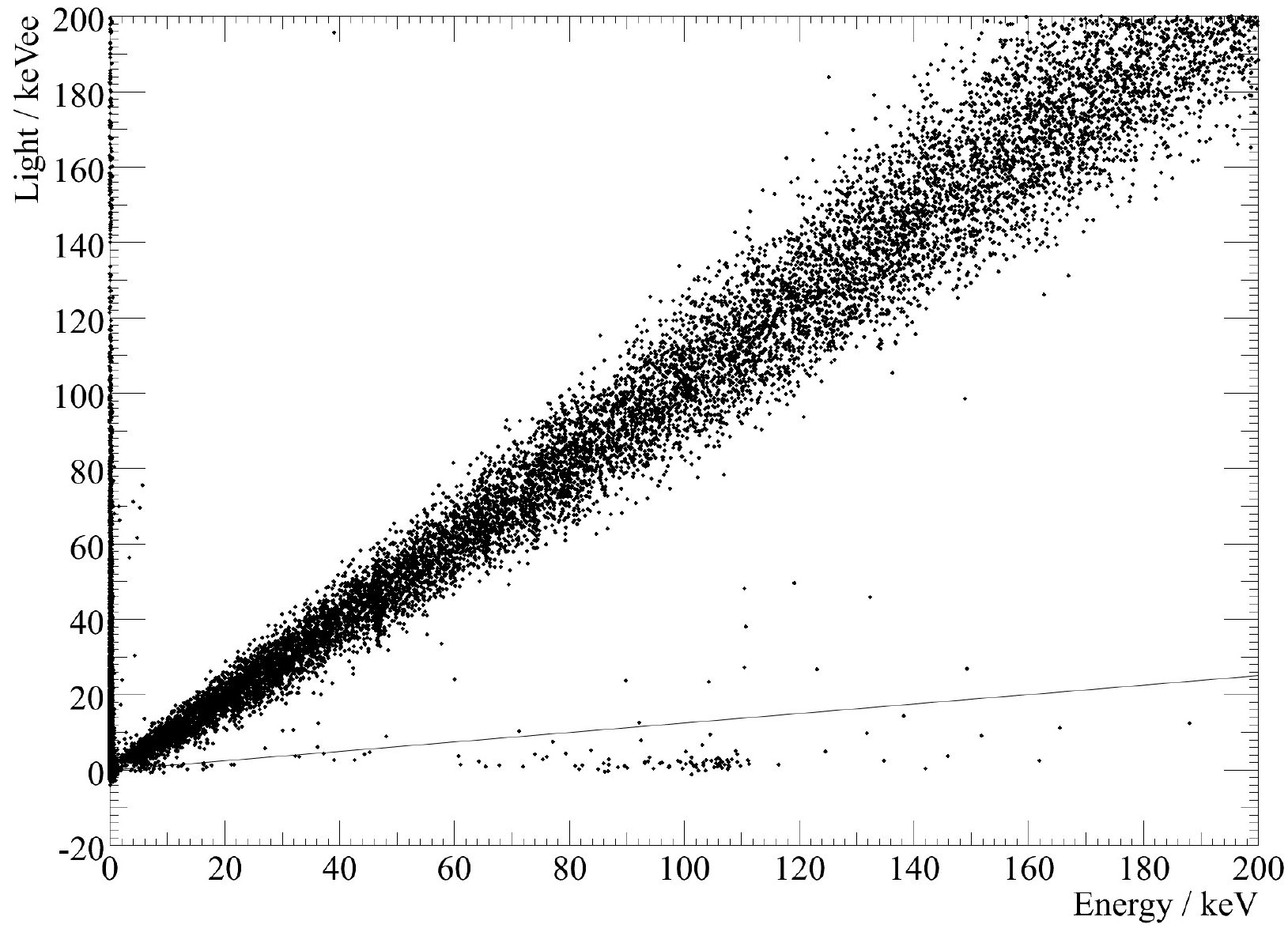}\end{center}
\vspace{-5mm}
\caption{Events observed in one crystal encapsulated within a silver reflector, over a period of about one month ($12.31\1{kg\,d}$, detector \textsc{Daisy}/run~27). We consider events below the line as nuclear recoils.}
\label{fig:DaisyRun27w}\end{figure}

Figure~\ref{fig:DaisyRun27w} shows a scatter plot of events recorded by one crystal enclosed in a silver reflector. The exposure is $12.31\1{kg\,d}$; no calibration source was present during this data taking. The horizontal axis shows the energy deposited in the phonon channel. The vertical axis shows energy measured by the light detector in units of keV electron equivalent $(\n{keV_{ee}})$, defined such that for $122\1{keV}$ gamma events (from a $\n{{}^{57}Co}$ calibration source) in the $\n{CaWO_4}$ crystal we register $122\1{keV_{ee}}$ in the light detector.

The diagonal band consists of electron recoils and gamma induced events. Nuclear recoil events yield significantly less light than electron or gamma events. Here, we define nuclear recoils to mean events below a slope of $25\1{keV_{ee}}/200\1{keV}$. Particles hitting the light detector directly show up along the vertical axis.

\begin{figure}[htbp]
\begin{center}\includegraphics[width=1\columnwidth,clip,trim=0 0 47 140]{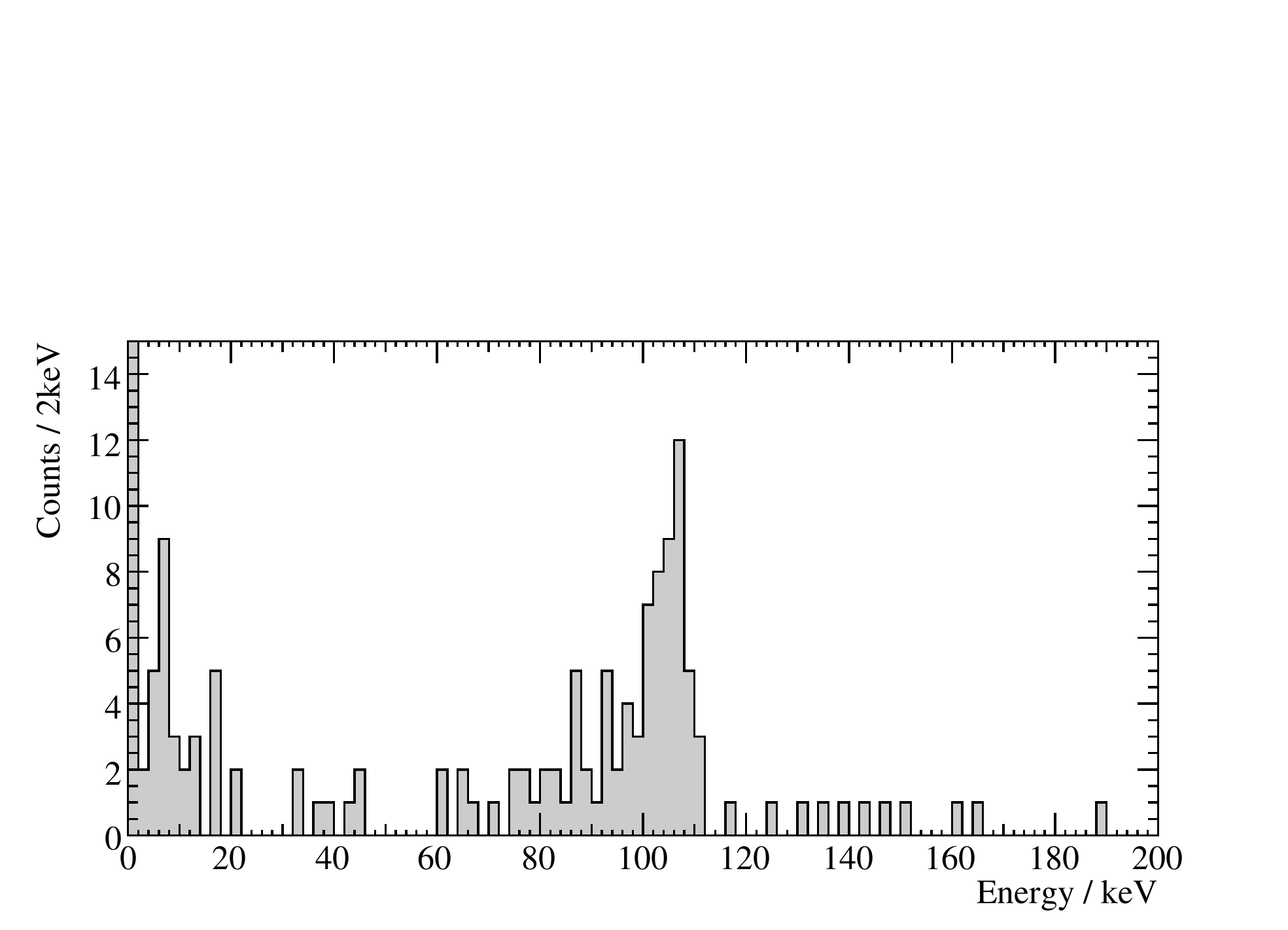}\end{center}
\vspace{-5mm}
\caption{Spectrum of events in the nuclear recoil band of figure~\ref{fig:DaisyRun27w}. The prominent feature around $103\1{keV}$ is attributed to the decay of $\n{{}^{210}Po}$ on and below nearby surfaces. The continuous background is attributed to neutrons, since at the time of data taking no neutron shield was present yet.}
\label{fig:SpectrumRun27e}
\end{figure}

The spectrum of nuclear recoil events is shown in figure~\ref{fig:SpectrumRun27e}. The continuum of events in this band is dominated by neutron induced events, since at the time of data taking, the detector was not yet surrounded with a neutron shield. In addition, a population around $\sim100\1{keV}$ is visible which we attribute to recoiling $\n{{}^{206}Pb}$ nuclei following the decay of $\n{{}^{210}Po}$ in the vicinity of the crystals. If the $\n{{}^{210}Po}$ decay happens on a nearby surface, decays can result in lead nuclei impinging on the target crystal. From simple momentum conservation the energy of these nuclei is fixed to $103\1{keV}$. Furthermore, if $\n{{}^{210}Po}$ is implanted in the crystal very close to its surface, the alpha particle may also deposit part of its energy in the target. Then our calorimetric measurement results in an energy higher than $103\1{keV}$. Finally, a more relevant possibility occurs if the polonium is implanted in nearby surfaces such as the reflective foil. Then the lead nucleus loses energy on its way to the crystal, resulting in an event with less than $103\1{keV}$. This can be a dangerous background to direct Dark Matter searches. With our $\n{CaWO_4}$ crystals in particular, such lead recoils mimic Dark Matter induced tungsten recoils, which are expected in the energy region below a few tens of keV~\cite{westphal2008a,angloher2009}. It is therefore indispensable to be able to distinguish this background from tungsten recoils.

\begin{figure}[htbp]
\begin{center}\includegraphics[width=1\columnwidth,clip,trim=0 0 50 140]{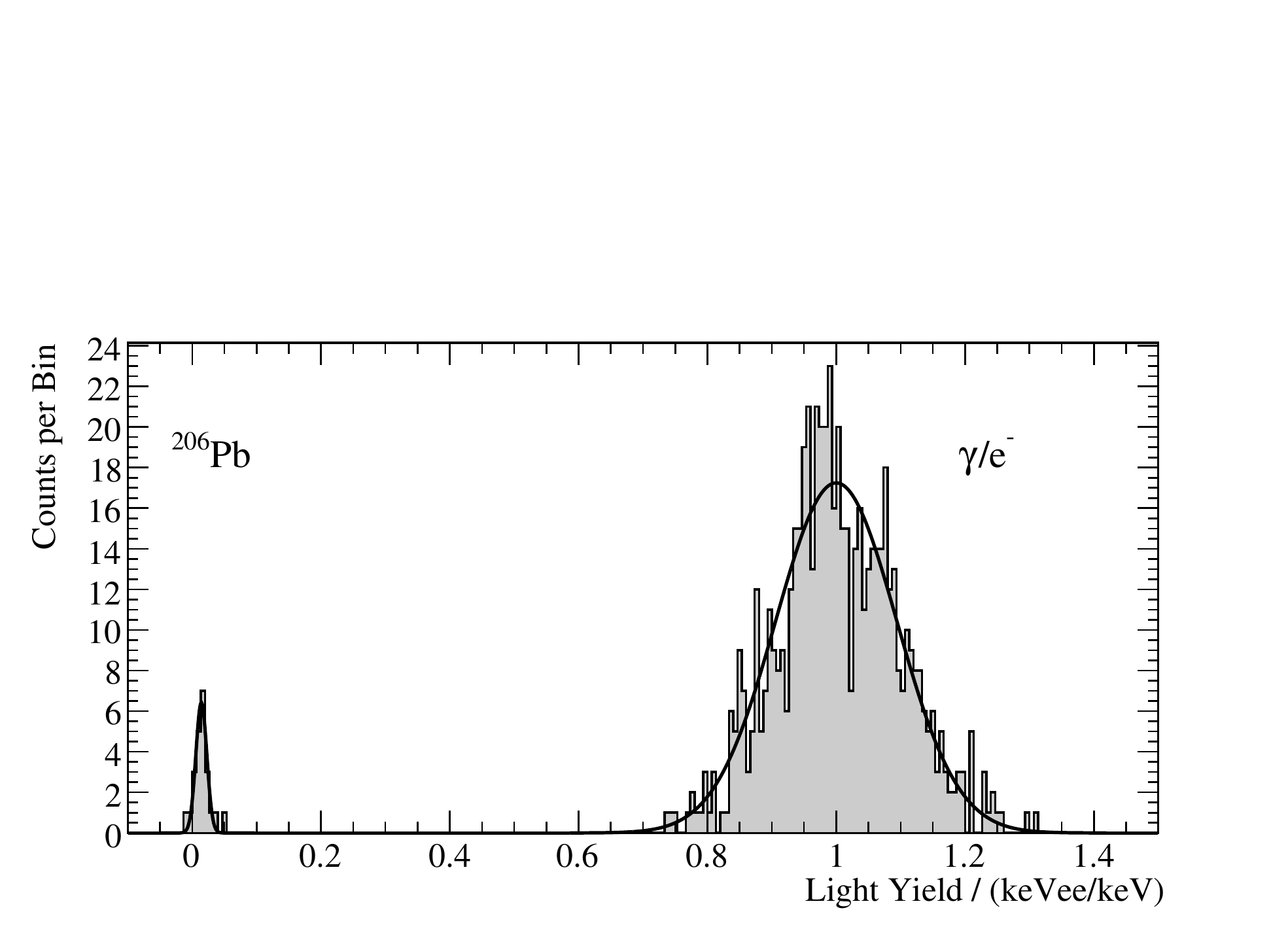}\end{center}
\vspace{-5mm}\caption{Light Yield of events between $93\1{keV}$ and $103\1{keV}$. The right population is a cross section of the gamma and electron recoil band with a light yield of unity. The left population comes from $\n{{}^{206}Pb}$ nuclei hitting the crystal. A Gaussian fit gives the light yield of these events as $(0.0142\pm0.0013)\1{keV_{ee}/keV}$.}
\label{fig:PbLightYieldE}\end{figure}

Having identified the population as being due to recoiling lead nuclei allows us to measure the amount of light emitted by these events. This quantity is also relevant to the search for Dark Matter due to the similarity of recoiling lead nuclei and the tungsten recoils that are expected from Dark Matter. We define the light yield as energy in the light detector (in $\n{keV_{ee}}$) over energy in the phonon detector (in $\n{keV}$). To measure its value for lead recoils, one has to exclude events above $103\1{keV}$ which possibly have a light contribution from the alpha particle. As a lower limit we use $93\1{keV}$ to keep the influence of neutrons small. Figure~\ref{fig:PbLightYieldE} shows the light yield of these events. From a Gaussian fit the light yield of $\sim100\1{keV}$ lead recoils at cryogenic temperatures of about $15\1{mK}$ in $\n{CaWO_4}$ crystals is found to be $(0.0142\pm0.0013)$ compared to that of electrons of the same energy. This value is robust within the stated errors when small variations of the chosen energy interval are included.

\section{Po-210 Decay Observed with Scintillating Reflectors}

\begin{figure}[htbp]
\begin{center}\includegraphics[width=1\columnwidth]{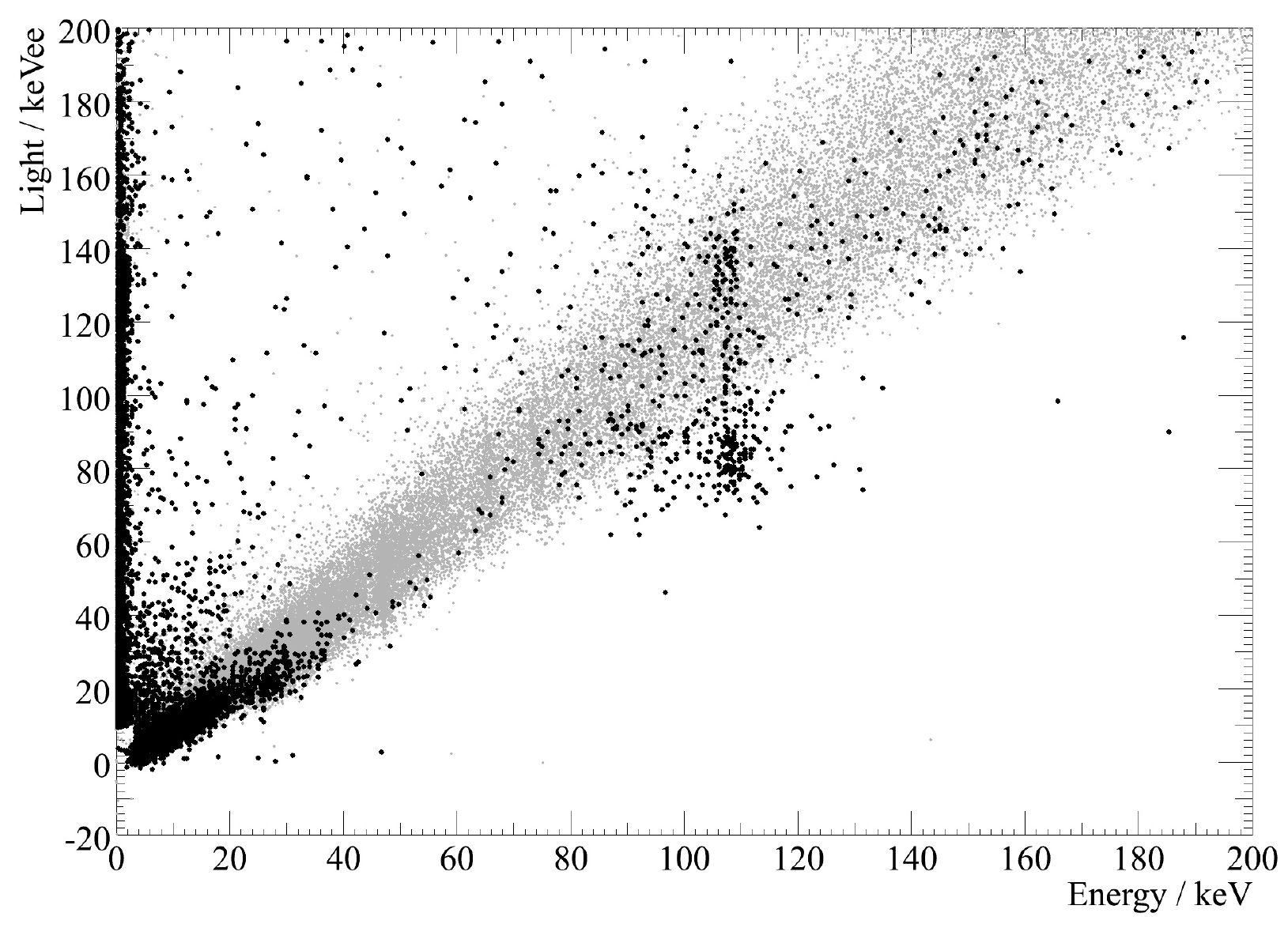}\end{center}
\vspace{-5mm}
\caption{Events observed in one detector module enclosed within a scintillating reflector (light dots, detector \textsc{Zora}/run~30). The exposure is $21.94\1{kg\,d}$, about two months of data taking. Events tagged by the selection criterion based on the pulse shape of the light detector are marked as heavy dots (see text for details).}
\label{fig:ZoraRun30u}\end{figure}

To distinguish this polonium induced background from Dark Matter induced tungsten recoils, we surround our detectors with a VM2000 class polymeric foil (from 3M), which is highly reflective and in addition also scintillates. Then the alpha particle from the polonium decay produces some additional scintillation light in the foil, shifting those events out of the region of nuclear recoils. Indeed, once this was done, the class of events observed in figure~\ref{fig:DaisyRun27w} disappeared from the nuclear recoil band. Figure~\ref{fig:ZoraRun30u} shows the corresponding scatter plot. The data shown is from another detector and a later data taking period, in which a neutron shield was present~\cite{angloher2009}, thus also eliminating the continuous neutron induced background in the nuclear recoil band.

The foil is highly efficient, and no events are seen in the nuclear recoil band around $100\1{keV}$. Comparing this scatter plot to figure~\ref{fig:DaisyRun27w}, we note that the scintillating foil causes a population of excess light events above the main gamma/electron band, dominantly at low energies in the phonon detector.

\begin{figure}[htbp]
\begin{center}\includegraphics[width=1\columnwidth,clip,trim=0 0 50 140]{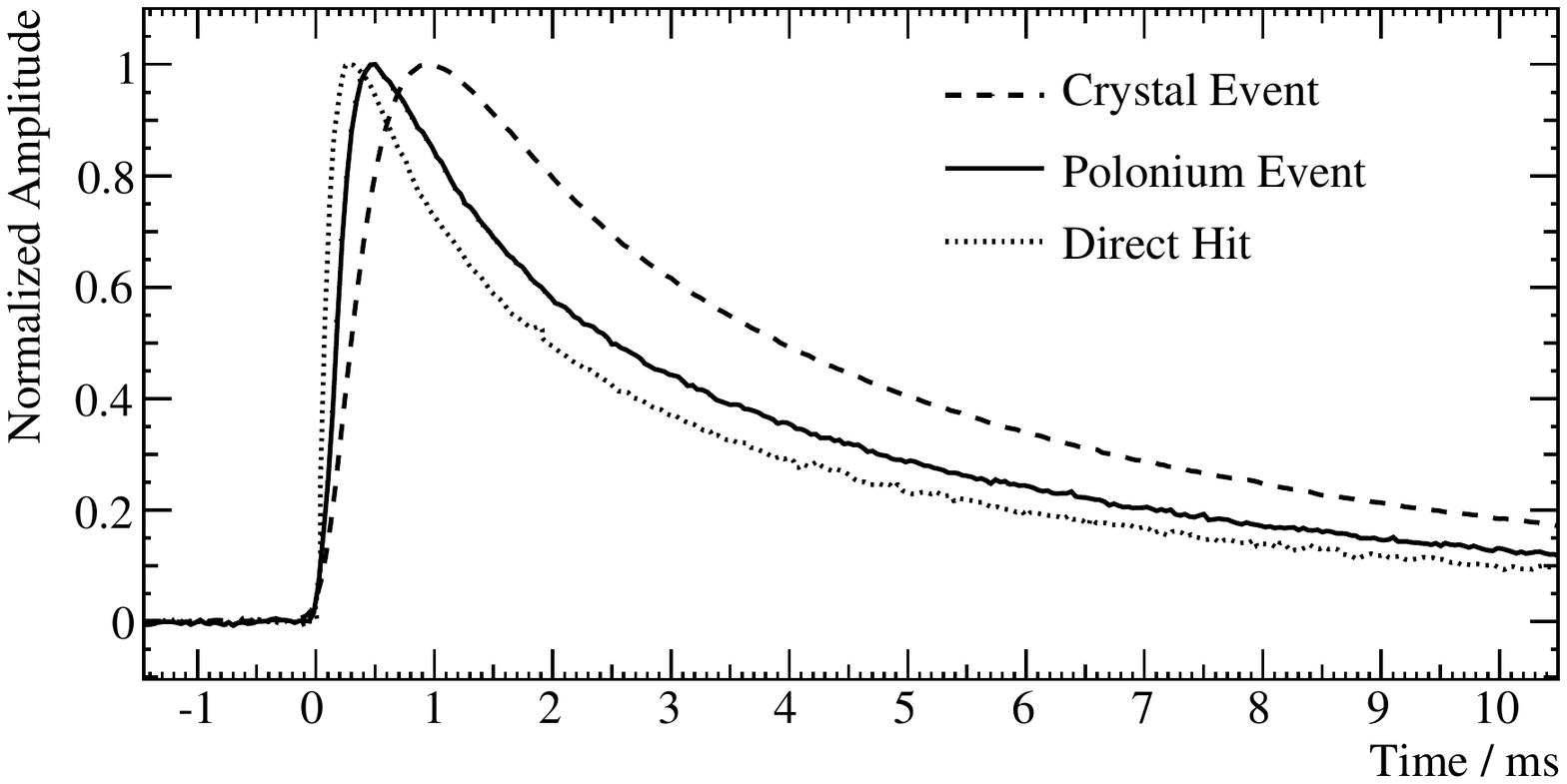}\end{center}
\vspace{-5mm}
\caption{Template events from the light detector, built from a few events of each class with energies of about $100\1{keV_{\n{ee}}}$: The fastest pulse is a direct hit in the light detector which takes only $\tau_{\n{rise}}=0.16\1{ms}$ from 10 to $90\percent$ peak height. A usual pulse from electron/gamma interactions in the crystal takes $\tau_{\n{rise}}=0.54\1{ms}$. A template of events from the $\n{{}^{210}Po}$ surface decays is also shown, seen to be in between the two other templates, with $\tau_{\n{rise}}=0.28\1{ms}$ (uncertainties are less than $0.01\1{ms}$).}
\label{fig:SOS21Shapes}\end{figure}

Incidentally, the polonium events are shifted to the gamma/electron recoil band. This can be illustrated by examination of the pulse shapes in the light detector. $\n{CaWO_4}$ is a rather slow scintillator in contrast to common plastic scintillators at millikelvin temperatures~\cite{distefano2003}. Hence pulses with an additional contribution from the scintillating foil are faster than pulses that are only due to an interaction in the crystal. This can be seen even by eye, comparing polonium events to crystal events in figure~\ref{fig:SOS21Shapes}.

Events from direct interactions in the light detector deposit no energy in the phonon detector and hence show up along the vertical axis in figure~\ref{fig:ZoraRun30u}. Such events deposit their energy instantaneously in the light detector and allow us to measure the detector response. This represents the fastest pulses possible, with much faster rise times than the usual events originating in the crystal, as can be seen in figure~\ref{fig:SOS21Shapes}. 

\begin{figure}[htbp]
\begin{center}\includegraphics[width=0.8\columnwidth,clip,trim=0 0 0 40]{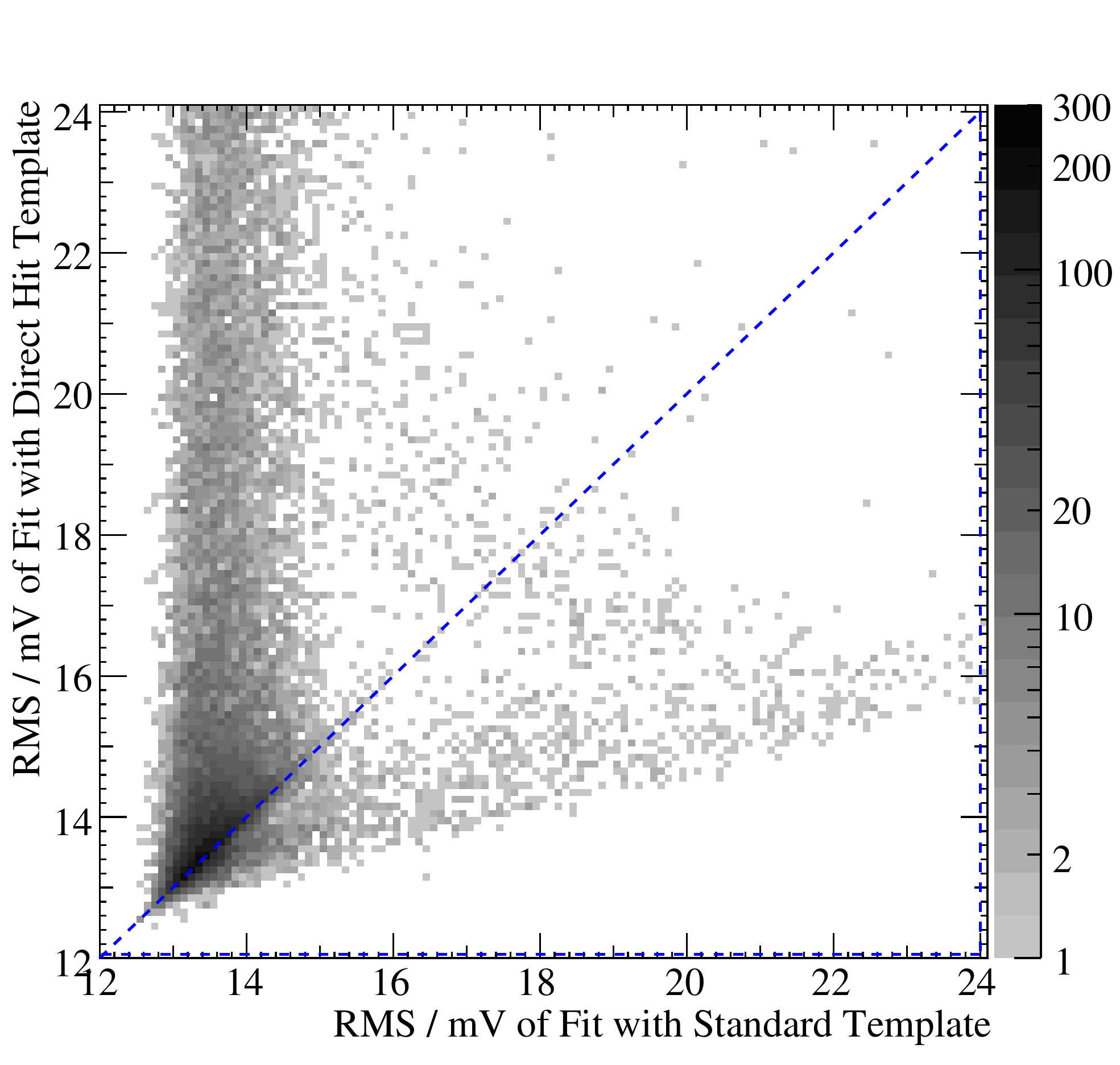}\end{center}
\vspace{-5mm}\caption{Histogram of events in figure~\ref{fig:ZoraRun30u} in the plane spanned by the two fit quality parameters. Entries per bin are color coded according to the logarithmic scale on the right. Events where the direct-hit template fits better (below the angle bisector) are tagged and shown as dark points in figure~\ref{fig:ZoraRun30u}.}
\label{fig:RMSTagging}\end{figure}

A second template pulse is built from such direct-hit events, and all events are fitted with both the standard template and this second template pulse. The qualities of the two individual template fits (their RMS values) are plotted against each other as in figure~\ref{fig:RMSTagging}. Events where the fast (direct hit) template pulse fits better are tagged as events which potentially have an additional light contribution from the scintillating foil. 

\begin{figure}[htbp]
\begin{center}\includegraphics[width=1\columnwidth,clip,trim=0 0 0 130]{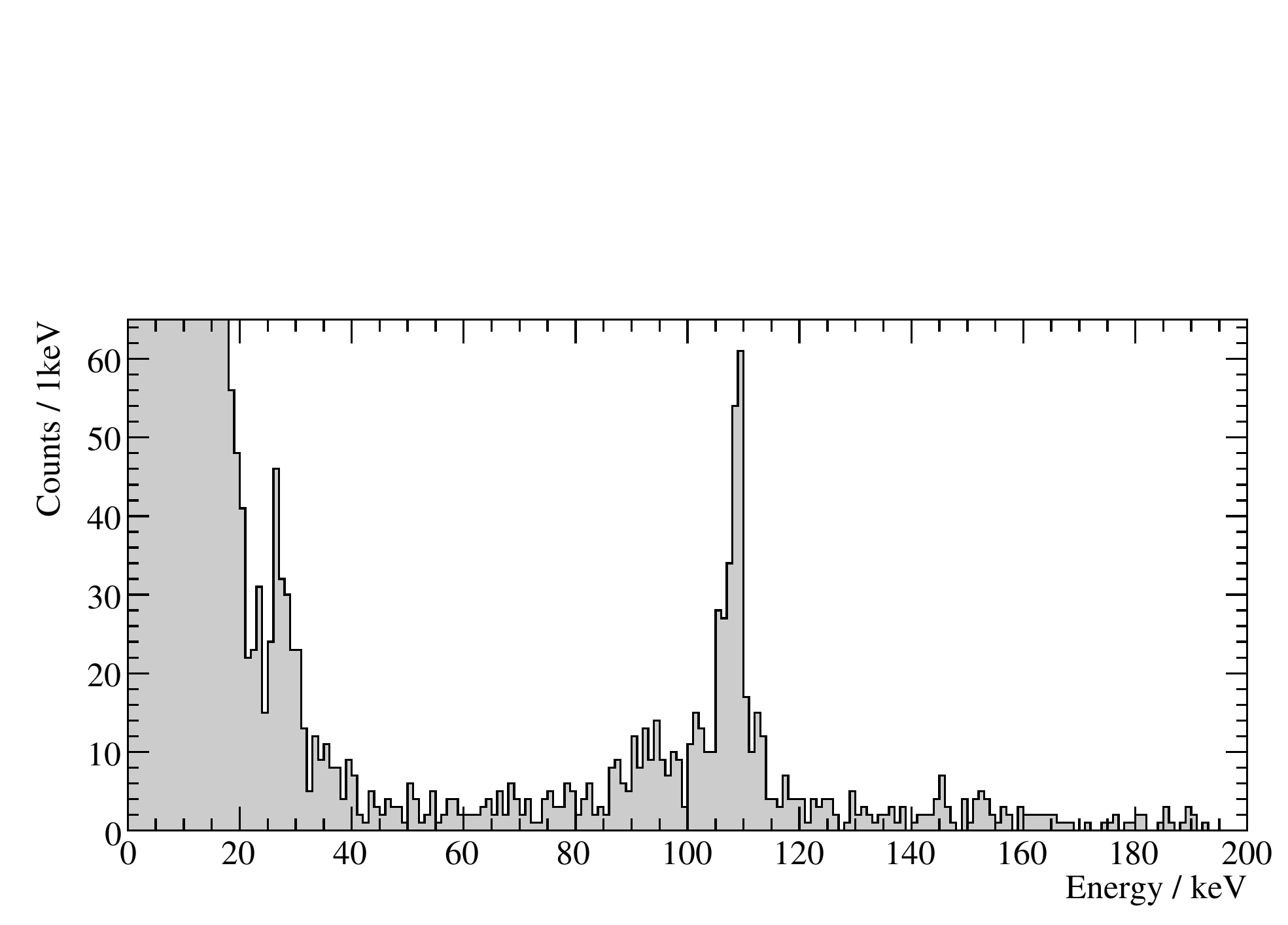}\end{center}
\vspace{-5mm}
\caption{Spectrum of events which are tagged based on the quality of the two template fits of the light detector. These events are shown as dark points in figure~\ref{fig:ZoraRun30u}. In contrast to figure~\ref{fig:SpectrumRun27e}, the detector was surrounded by a neutron shield, and the continuous background here is not due to neutrons, but due to leakage from the main gamma/electron band when applying the tagging procedure.}
\label{fig:SpectrumRun30s}
\end{figure}

In the scatter plot of figure~\ref{fig:ZoraRun30u} these events are highlighted, and their spectrum is shown in figure~\ref{fig:SpectrumRun30s}. Events of various origins are extracted by this tagging procedure: Particles hitting the light detector directly (they appear along the vertical axis in figure~\ref{fig:ZoraRun30u}) are recognized as such. As can be expected from pulse shape discrimination, the tagging procedure fails towards low energies in the light detector. This results in many mis-tagged events below $\sim40\1{keV_{ee}}$, where only less than a few $100\1{eV}$ are deposited in the light detector. Events leaking from the gamma/electron band form a continuous background in figure~\ref{fig:SpectrumRun30s}. Excess light events are also tagged, supporting our idea of their origin involving the faster scintillating foil. Finally, the polonium decay events are clearly identified and can be recovered from the gamma/electron band by this procedure.

\section{Pulse Shape of Lead Recoils}

\begin{figure}[htbp]
\begin{center}\includegraphics[width=1\columnwidth,clip,trim=0 0 0 160]{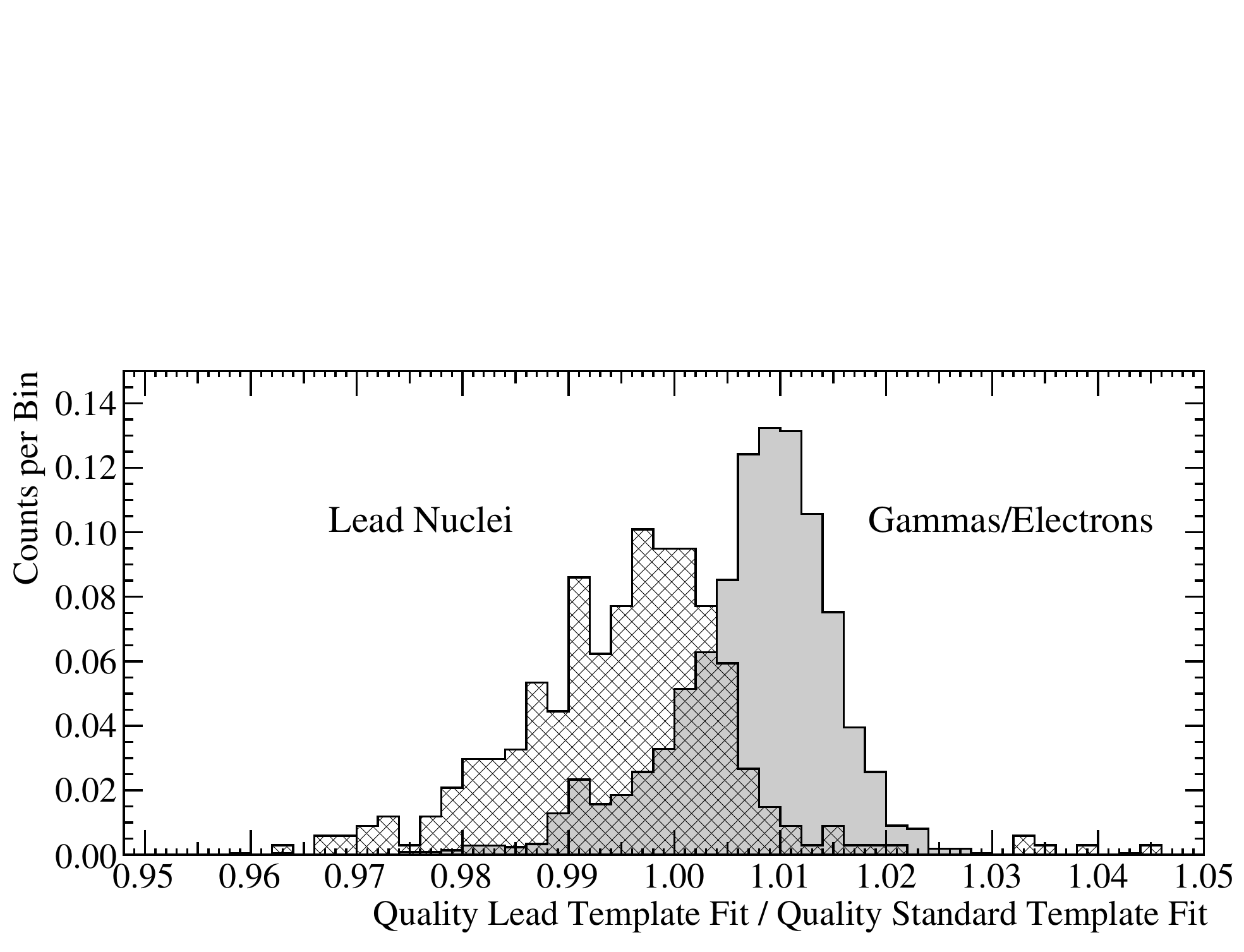}\end{center}
\vspace{-5mm}
\caption{Difference in phonon detector pulse shape for electron and nuclear recoils of about $100\1{keV}$. This difference is characterized by the ratio of the fit-RMS values obtained from lead template and standard template fits. Lead recoils were tagged by the different pulse shape in the light detector, which is possible in this particular case due to the additional light from the scintillating foil. Both distributions are normalized to unity to ease the comparison.}
\label{fig:LeadShape2}\end{figure}

Having extracted the events from the polonium decay based on the pulse shape in the light detector, we can examine the pulse shape in the phonon detector as well. By eye or based on simple parameters like the rise or decay times, no difference between the individual pulses can be seen. Yet a second template is built from these events around $103\1{keV}$ (figure~\ref{fig:SpectrumRun30s}). This again gives an additional fit quality parameter (the fit RMS), this time for the phonon detector. The ratio of this value and of the corresponding value of the standard template fit quality is plotted in figure~\ref{fig:LeadShape2}, for events in the energy interval $[100,120]\1{keV}$, both for the population of lead recoils from the polonium decay, and for standard electron or gamma events. The two populations clearly separate in this parameter on a level better than $1\sigma$. Hence the ratio of the quality of two standard event fits in the phonon detector can act as an additional discrimination parameter between electron and nuclear recoils.

\section{Conclusion}

The decay $\n{{}^{210}Po}\rightarrow\n{{}^{206}Pb}+\alpha$ results in recoiling lead nuclei, a potentially dangerous background in the search for Dark Matter. We find that in $\n{CaWO_4}$ at millikelvin temperatures, $100\1{keV}$ lead recoils have a light yield of $(0.0142\pm0.0013)$ compared to electrons of the same energy. In the case of scintillators as Dark Matter targets, we have shown a way to discriminate these lead recoils by surrounding the target with a scintillating foil, which leads to an enhanced light output for events from this background. Using the pulse shape of events in the light detector allows tagging of events which have a contribution from the scintillating foil for energies above $\sim 40\1{keV_{ee}}$. Thus the polonium induced lead recoils can be identified, and in addition, excess light events are tagged, revealing their origin as being due to the scintillating foil. Furthermore, we have shown that a difference exists in the pulse shape of the phonon-detector events for heavy nuclear recoils with respect to electron or gamma induced events at energies around $100\1{keV}$. This can be utilized to cross-check the origin of events in the phonon channel.

\section{Acknowledgments}

This work was partially supported by funds of the DFG (SFB 375, Transregio 27 ``Neutrinos and Beyond''), the Munich Cluster of Excellence (``Origin and Structure of the Universe"), the EU networks for Cryogenic Detectors (ERB-FMRXCT980167) and for Applied Cryogenic Detectors (HPRN-CT2002-00322), and the Maier-Leibnitz-Laboratorium (Garching). Support was provided by the Science and Technology Facilities Council.


\end{document}